\documentclass{report}
\usepackage{epsfig}
\renewcommand{\abstract}[1]{{\footnotesize \noindent {\bf Abstract} #1 \\}}
\renewcommand{\author}[1]{\subsection*{#1}}
\newcommand{\address}[1]{\subsection*{\it#1}}
\def\etal{{\em et al}}

\def\rme{{\rm e}}
\def\lessim{\mathrel{\rlap{\raise.5ex\hbox{$<$}}{\lower.5ex\hbox{$\sim$}}}}
\def\gtrsim{\mathrel{\rlap{\raise.5ex\hbox{$>$}}{\lower.5ex\hbox{$\sim$}}}}
\def\H{{\rm H}}
\def\D{{\rm D}}
\def\T{{\rm T}}
\def\he#1{\hbox{$^{#1}$He}}
\def\li#1{\hbox{$^{#1}$Li}}
\def\be#1{\hbox{$^{#1}$Be}}
\def\B{{\rm B}}

\def\CMB{{\rm CMB}}
\def\F{{\rm F}}
\def\fr{{\rm fr}}
\def\hii{H~{\rm II}}
\def\N{{\rm N}}
\def\np{{n\,\leftrightarrow\,p}}
\def\p{{\rm p}}
\def\popii{Pop~{\rm II}}

\def\yp{\hbox{$Y_{\rm p}$}}
\begin{document}

\chapter*{Measuring the baryon content of the universe: BBN vs CMB}
\author{Subir Sarkar}
\address{Dept of Physics, Oxford University, 1 Keble Road, 
 Oxford OX1 3NP, UK}

\abstract{The relic abundance of baryons --- the only form of stable
matter whose existence we are certain of --- is a crucial parameter
for many cosmological processes, as well as material evidence that
there is new physics beyond the Standard Model. We discuss recent
determinations of the cosmological baryon density from analysis of the
abundances of light elements synthesised at the end of ``the first
three minutes'', and from the observed temperature anisotropies
imprinted on small angular-scales in the cosmic microwave background
when the universe was $\sim10^5$ yr old.}

\section{Introduction} 

Begining as an uniformly distributed plasma in the radiation-dominated
era, baryons are now distributed in a hierarchy of structures, from
galaxies to superclusters, which have formed through gravitational
collapse and constitute the familiar visible universe. Although it is
recognised that baryons are a dynamically unimportant component,
outweighed by the dark matter that actually holds such structures
together, they constitute the only form of stable matter which we know
about and can study directly. Nevertheless their cosmological origin
is a mystery, even harder to fathom than that of the much more
dominant dark matter. This is not always appreciated, particularly by
those who think of baryons as familiar `ordinary' stuff, as opposed to
the `exotic' particles that particle theorists dream up such as
supersymmetric neutralinos. However it follows from elementary kinetic
theory \cite{relic} that the relic abundance of massive particles (and
anti-particles) which were in thermal equilibrium in the early
universe is proportional to the inverse of their (velocity-averaged)
self-annihilation cross-section: $\Omega_{x}h^2\sim3\times10^{-10}{\rm
GeV}^{-2}/\langle\sigma\,v\rangle_{x\bar{x}}$. Thus strongly
interacting particles such as baryons should barely have survived
self-annihilations
($\Omega_{\B}h^2\sim10^{-11}\Rightarrow\,n_\B/n_\gamma\sim10^{-18}$),
while weakly interacting particles such as stable neutralinos, if they
exist, should naturally have a present density which is dynamically
important ($\Omega_{\chi}h^2\sim0.1$). Since $n_\B/n_\gamma$ is
actually found to be of ${\cal O}(10^{-10})$ today, with no evidence
for antimatter, it is thus {\em necessary} to postulate that there was
an initial excess of quarks over anti-quarks by about 1 part in
$10^{9}$, before baryons and anti-baryons first formed following the
QCD confinement transition at $\sim0.2$~GeV and began to
annihilate. As Sakharov first noted, to create this baryon asymmetry
of the universe (BAU) requires new physics, specifically the violation
of baryon number, as well as of $C$ and $CP$, together with departure
from thermal equilibrium to provide an arrow of time. Whereas all
these ingredients may in principle be available in the Standard
$SU(3){\otimes}SU(2){\otimes}U(1)$ Model (SM) cosmology \cite{bausm},
in practice it has not proved possible to generate the required BAU
with SM dynamics, essentially because the LEP bound on the Higgs mass
precludes a strong enough first-order electroweak symmetry breaking
phase transition. There is still hope that this may prove possible in
supersymmetric extensions of the SM, which moreover may have new
sources of $CP$ violation \cite{baususy}. There is also the novel
possibility that the BAU may be linked to the smallness of neutrino
masses since the source for both of them may be lepton number
violating dynamics at energies close to the GUT scale
\cite{baulepto}. Of course the source of the BAU may be some totally
different mechanism, e.g. the Affleck-Dine mechanism in supergravity
\cite{baurev}, and there is no guarantee that we will necessarily ever
be able to link it directly to laboratory physics. All this makes it
quite clear that the existence of `ordinary' matter today is far more
mysterious in principle than that of dark `exotic' matter. Our very
existence {\em requires} that there is exciting new physics to be
discovered beyond the Standard Model!

My task here is to review recent determinations of the baryon density
of the universe which is not only of fundamental significance as
discussed above, but also an important parameter for crucial
cosmological processes, in particular for Big Bang nucleosynthesis
(BBN) of the light elements at $t\sim10^2$~s \cite{bbnhist}, and for
the decoupling of the cosmic microwave background (CMB) when the
universe turns neutral at $t\sim10^{5}$~yr \cite{cmbhist}. It has been
recognised for some years that the primordial abundance of a fragile
element such as deuterium, whose only source is BBN, serves as a
sensitive probe of the baryon density \cite{rafs}. More recently it
has been noted that the temperature fluctuations imprinted on the CMB
at small angular scales by acoustic oscillations of the coupled
photon-baryon plasma during (re)combination \cite{hs} enable an
independent determination of the baryon density from CMB sky maps
\cite{jkks}. Recent observations of light element abundances
(particularly the deuterium abundance in quasar absorption systems at
high redshift), as well as of sub-degree scale temperature
fluctuations in the CMB, have allowed a comparison of these two
independent methods. After some initial problems, the two
determinations are now believed to be in good agreement, and moreover
consistent with estimates of the baryon content of the high redshift
Lyman-$\alpha$ `forest' \cite{lyalpha}, leading some cosmologists to
declare this a triumph for ``precision cosmology''. This may however
be premature because there still remain major observational
uncertainties on the BBN side, while cosmological parameter extraction
from the CMB requires important assumptions, particularly concerning
{\em the nature and spectral shape of the primordial density
perturbations}. It is undoubtedly impressive that the standard
cosmological model \cite{relic} (extended to include the standard
model of structure formation from scale-free adiabatic initial density
perturbations) has passed an important consistency check. However
since cosmologists are no longer starved of data, it would seem more
appropriate to abandon the untested assumptions of the model and begin
to confront a more sophisticated paradigm, rather than exult in having
achieved agreement between different model-dependent determinations to
within a factor of $\sim2$. I will therefore emphasise the loopholes
in the present approaches to determination of the baryon density in
the hope that this will motivate observers to rule out such
``non-standard'' possibilities. Only then would we really be justified
in calling it the `standard model' of cosmology.

\section{Big bang nucleosynthesis and the baryon density}

There have been many discussions of how the baryon density influences
the synthesis of the light elements \cite{bbnrev}; here we follow a
recent summary \cite{pdg}. BBN is sensitive to physical conditions in
the early radiation-dominated era at temperatures $T\lessim1$~MeV,
corresponding to an age $\gtrsim1$~s. At higher temperatures, weak
interactions were in thermal equilibrium, thus fixing the ratio of the
neutron and proton number densities to be $n/p=\rme^{-Q/T}$, where
$Q=1.293$~MeV is the neutron-proton mass difference. As the
temperature dropped, the neutron-proton inter-conversion rate,
$\Gamma_{\np}\sim\,G_\F^2T^5$, fell faster than the Hubble expansion
rate, $H\sim\sqrt{g_*G_\N}\;T^2$, where $g_*$ counts the number of
relativistic particle species determining the energy density in
radiation. This resulted in breaking of chemical equilibrium
(``freeze-out'') at $T_\fr\sim(g_*G_\N/G_\F^4)^{1/6}\simeq1$~MeV. The
neutron fraction at this time, $n/p=\rme^{-Q/T_\fr}\simeq1/6$ is thus
sensitive to every known physical interaction, since $Q$ is determined
by both strong and electromagnetic interactions while $T_\fr$ depends
on the weak as well as gravitational interactions. Moreover it is
sensitive to the Hubble expansion rate, affording e.g. a probe of the
number of relativistic neutrino species. After freeze-out the neutrons
were free to $\beta$-decay so the neutron fraction dropped to
$\simeq1/7$ by the time nuclear reactions began. The rates of these
reactions depend on the density of baryons (strictly speaking, {\em
nucleons}), which is usually normalised to the blackbody photon
density as $\eta\equiv\,n_\B/n_\gamma$. As we shall see, all the
light-element abundances can be explained with
$\eta_{10}\equiv\eta\div10^{-10}$ in the range 2.6--6.2.

The nucleosynthesis chain begins with the formation of deuterium in
the process $p(n,\gamma)\D$. However, photo-dissociation by the high
number density of photons delays production of deuterium (and other
complex nuclei) until well after $T$ drops below the binding energy of
deuterium, $\Delta_\D=2.23$~MeV. The quantity
$\eta^{-1}\rme^{-\Delta_\D/T}$, i.e. the number of photons per baryon
above the deuterium photo-dissociation threshold, falls below unity at
$T\simeq0.1$~MeV; nuclei can then begin to form without being
immediately photo-dissociated again. Only 2-body reactions such as
$\D(p,\gamma)\he3$, $\he3(\D,p)\he4$, are important because the
density is rather low at this time --- about the density of water!
Nearly all the surviving neutrons when nucleosynthesis begins end up
bound in the most stable light element \he4. Heavier nuclei do not
form in any significant quantity both because of the absence of stable
nuclei with mass number 5 or 8 (which impedes nucleosynthesis via
$n\he4$, $p\he4$ or $\he4\he4$ reactions) and the large Coulomb
barriers for reactions such as $\T(\he4,\gamma)\li7$ and
$\he3(\he4,\gamma)\be7$. Hence the primordial mass fraction of \he4,
conventionally referred to as $Y_\p$, can be estimated by the simple
counting argument
\begin{equation}
 Y_\p = {2(n/p) \over 1+n/p} \simeq 0.25 .
\label{Yp}
\end{equation}
There is little sensitivity here to the actual nuclear reaction rates,
which are however important in determining the other ``left-over''
abundances: D and \he3 at the level of a few times $10^{-5}$ by number
relative to H, and \li7/H at the level of about $10^{-10}$. The
experimental parameter most important in determining $Y_\p$ is the
neutron lifetime, $\tau_n$, which normalises (the inverse of)
$\Gamma_{\np}$. The experimental uncertainty in $\tau_n$ used to be a
source of concern but has recently been reduced substantially:
$\tau_n=885.7\pm0.8$~s.

\begin{figure}[htb]
\centerline{\epsfxsize6.5cm\epsffile{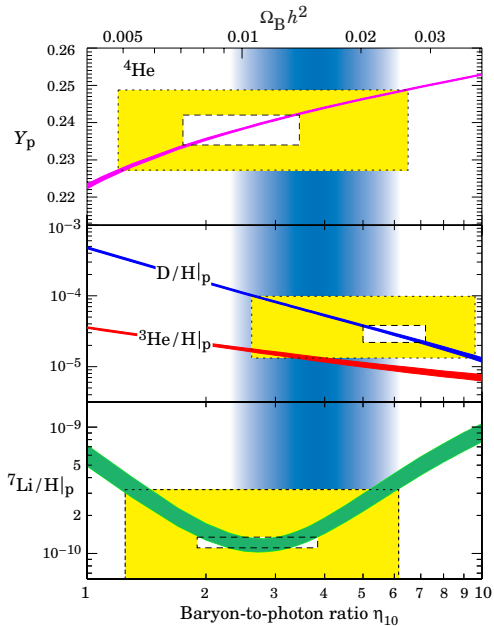}}
\bigskip
\caption{The primordial abundances of \he4, D, \he3 and \li7 as
predicted by the standard BBN model compared to observations ---
smaller boxes: $2\sigma$ statistical errors; larger boxes: $\pm
2\sigma$ statistical and systematic errors added in quadrature (from
Ref.\protect\cite{pdg}).}
\label{abundantlight}
\end{figure}

The predicted elemental abundances, calculated using the (publicly
available \cite{sark}) Wagoner code \cite{bbnhist,kaw}, are shown in
Fig.~\ref{abundantlight} as a function of $\eta_{10}$. The \he4 curve
includes small corrections due to radiative processes at zero and
finite temperature, non-equilibrium neutrino heating during $e^\pm$
annihilation, and finite nucleon mass effects \cite{lt,emmp}; the
range reflects primarily the 1$\sigma$ uncertainty in the neutron
lifetime. The spread in the curves for D, \he3 and \li7 corresponds to
the 1$\sigma$ uncertainties in nuclear cross sections estimated by
Monte Carlo methods \cite{skm}; polynomial fits to the predicted
abundances and the error correlation matrix have been given
\cite{flsv}. Recently the input nuclear data have been carefully
reassessed \cite{nb,cfo}, leading to improved precision in the
abundance predictions. The boxes in Fig.\ref{abundantlight} show the
observationally inferred primordial abundances with their associated
uncertainties, as discussed below.

\subsection{Primordial light element abundances} 

BBN theory predicts the universal abundances of D, \he3, \he4 and
\li7, which are essentially determined by $t\sim180$~s (although some
reactions continue for several hours). Abundances are however
observed at much later epochs, after stellar nucleosynthesis has
already commenced. The ejected remains of this stellar processing can
alter the light element abundances from their primordial values, but
also produce heavy elements such as C, N, O, and Fe (``metals''). Thus
one seeks astrophysical sites with low metal abundances, in order to
measure light element abundances which are closer to primordial.

\begin{figure}[htb]
\centerline{
\epsfysize5.6cm\epsffile{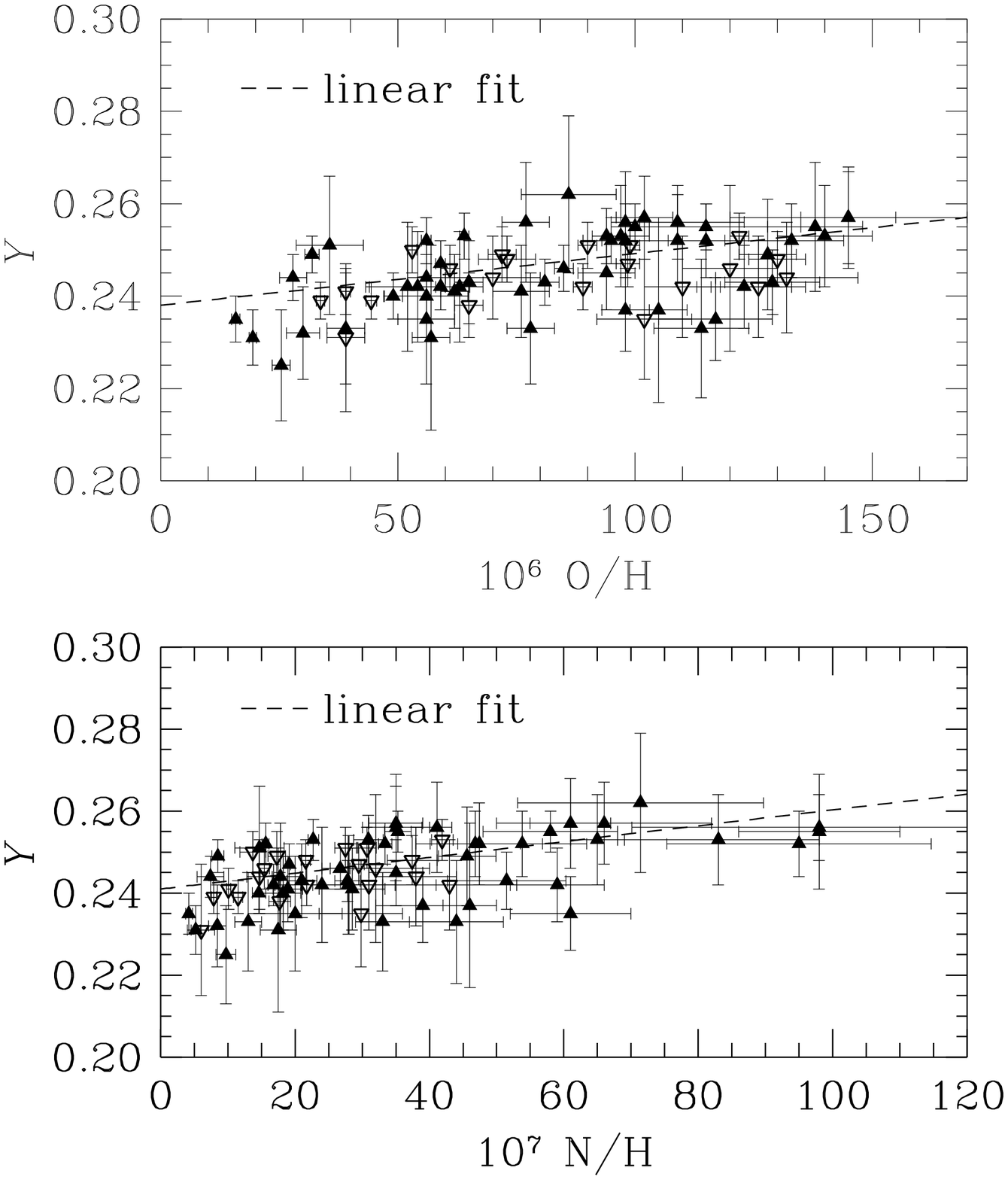}\hspace{1cm}
\epsfysize5.5cm\epsffile{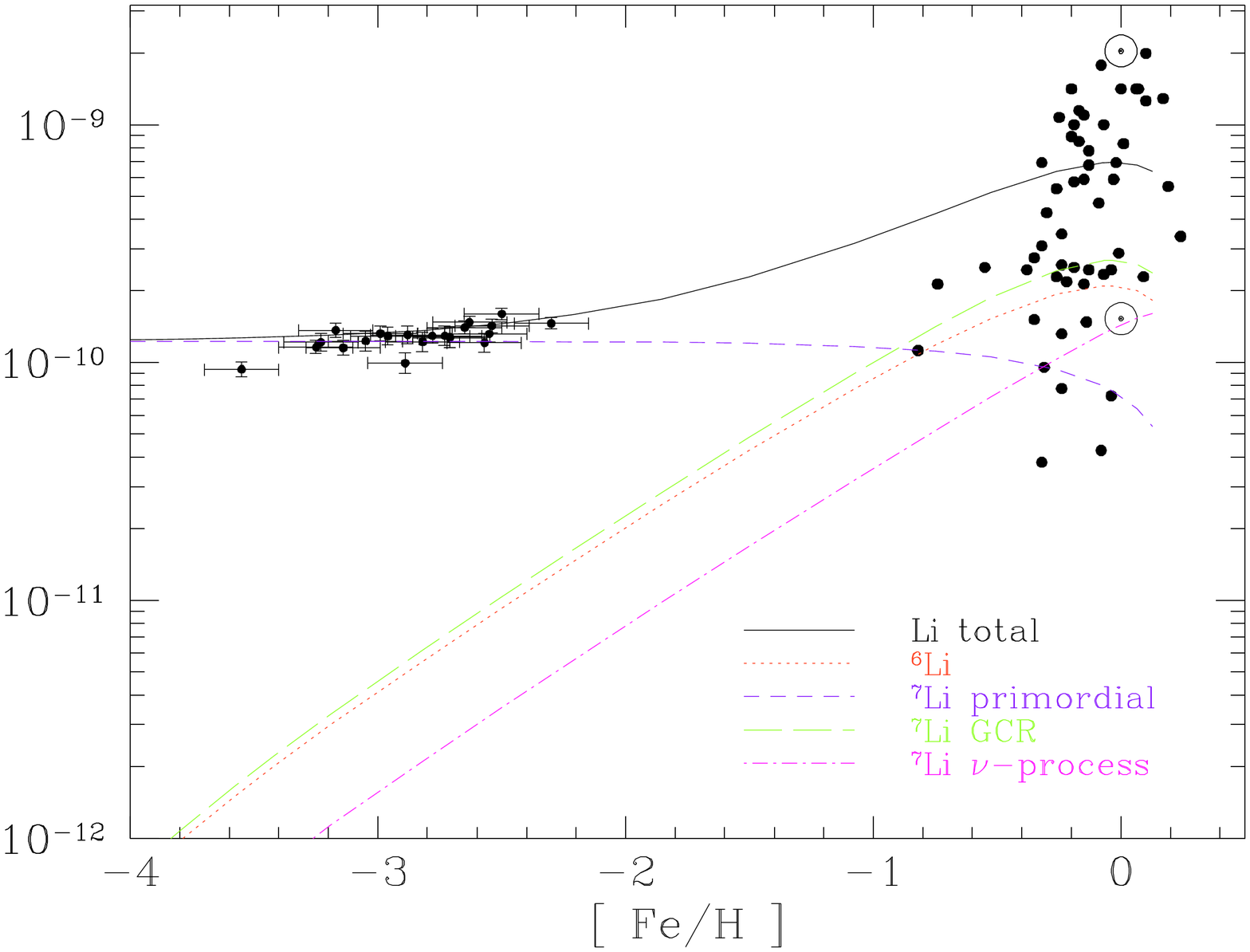}}
\bigskip
\caption{The primordial abundances of \he4 (from
Ref.\protect\cite{fo}) and of \li7 (from Ref.\protect\cite{rbofn}),
inferred from the observed abundances in, respectively, BCGs and
\popii\ stars.}
\label{He4Li7}
\end{figure}

We observe \he4 in clouds of ionized hydrogen (\hii\ regions), the
most metal-poor of which are in dwarf blue compact galaxies
(BCGs). There is now a large body of data on \he4 and C, N, O in these
systems \cite{pste,izo}. These data confirm that the small stellar
contribution to helium is positively correlated with metal production
(see Fig.\ref{He4Li7}); extrapolating to zero metallicity gives the
primordial \he4 abundance \cite{fo}:
\begin{equation}
 \yp = 0.238 \pm 0.002 \pm 0.005 \ . 
\label{YpObs} 
\end{equation}
Here and throughout, the first error is statistical, and the second is
an estimate of the systematic uncertainty. The latter clearly
dominates, and is based on the scatter in different analyses of the
physical properties of the \hii\ regions \cite{pste,izo,os,sj}. Other
extrapolations to zero metallicity give $\yp=0.244\pm0.002$
\cite{izo}, and $\yp=0.235\pm0.003$ \cite{ppr}, while the average in
the 5 most metal-poor objects is $\yp=0.238\pm0.003$ \cite{ppl}. The
value in Eq.(\ref{YpObs}), shown in Fig.\ref{abundantlight}, is
consistent with all these determinations.

The systems best suited for Li observations are hot, metal-poor stars
belonging to the halo population (\popii) of our Galaxy. Observations
have long shown that Li does not vary significantly in such stars
having metallicities $\lessim1/30$ of solar --- the ``Spite plateau''
\cite{ss,mpb}. Recent precision data suggest a small but significant
correlation between Li and Fe \cite{rnb} which can be understood as
the result of Li production from cosmic rays
\cite{sfosw}. Extrapolating to zero metallicity (see
Fig.\ref{He4Li7}) one arrives at a primordial value \cite{rbofn}
\begin{equation}
{\rm Li/H}|_\p =
 (1.23 \pm 0.06 {}^{+0.68}_{-0.32} {}^{+0.56}) \times 10^{-10} \ .
\label{Lip}
\end{equation}
The last error is our estimate of the maximum upward correction
necessary to allow for possible destruction of Li in \popii\ stars,
due to e.g. mixing of the outer layers with the hotter interior
\cite{ddk,vau}. Such processes can be constrained by the absence of
significant scatter in the Li-Fe correlation plot \cite{mpb,rnb}, and
through observations of the even more fragile isotope \li6
\cite{sfosw}.

\begin{figure}[htb]
\centerline{
\epsfysize5.5cm\epsffile{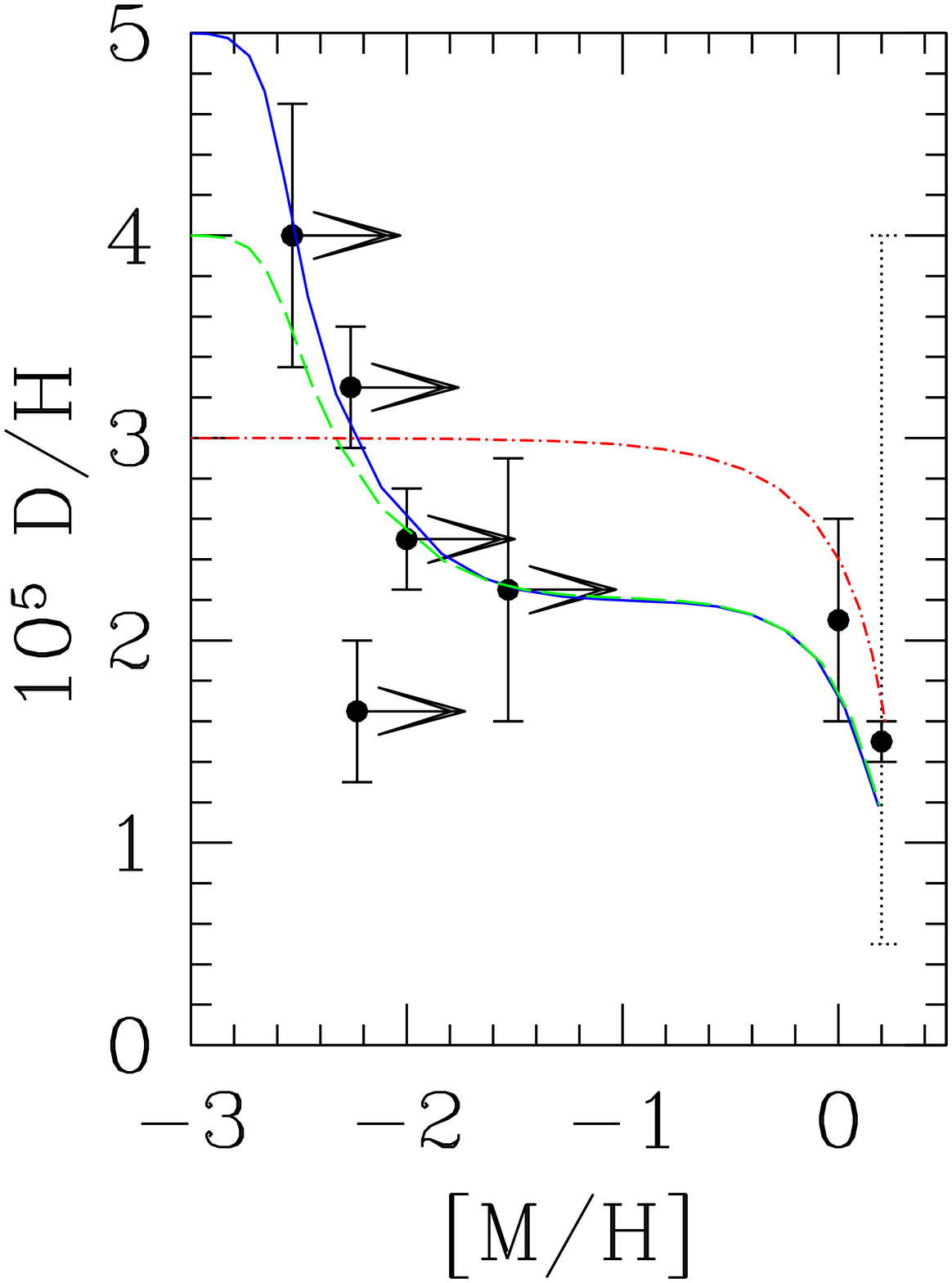}\hspace{1cm}
\epsfysize5cm\epsffile{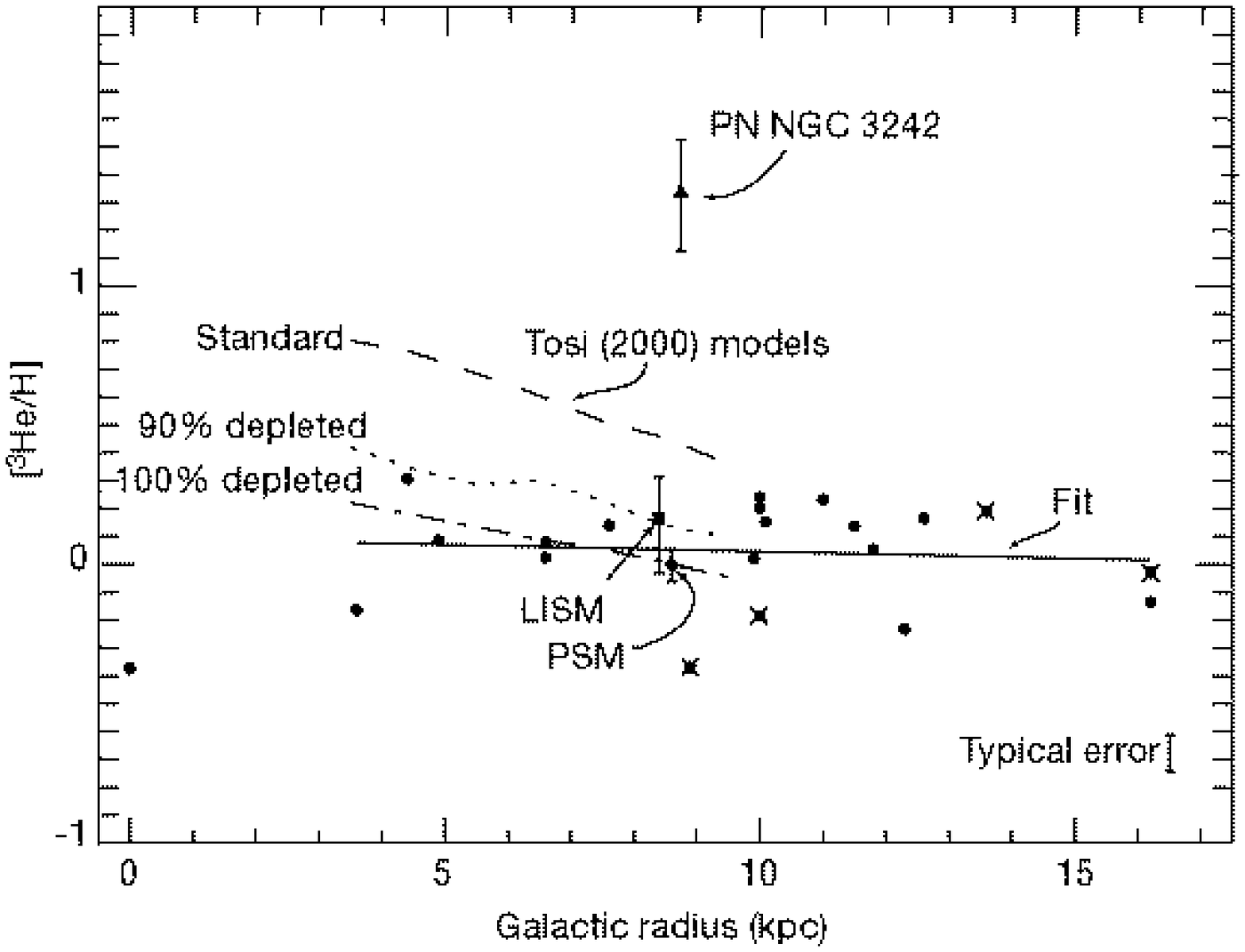}}
\bigskip
\caption{Chemical evolution models compared to the observed abundance
of \D\ in QAS and in the ISM (from Ref.\protect\cite{foscv}), and of
\he3 in galactic HII regions (from Ref.\protect\cite{bbrw}).}
\label{DHe3}
\end{figure}

In recent years, high-resolution spectra have revealed the presence of
D in quasar absorption systems (QAS) at high-redshift , via its
isotope-shifted Lyman-$\alpha$ absorption
\cite{schr,tfb,webb,ome,dod,pb}. It is believed that there are no
astrophysical sources of deuterium \cite{els}, so any measurement of
D/H provides a lower limit to the primordial abundance and thus an
upper limit on $\eta$; for example, the local interstellar value of
$\D/\H=(1.5\pm0.1)\times10^{-5}$ \cite{lin} requires
$\eta_{10}\le9$. Early reports of $\D/\H>10^{-4}$ towards 2 quasars
(Q0014+813 \cite{schr} and PG1718+4807 \cite{webb}) have been
undermined by later analyses \cite{bkt,kirk}. Three high quality
observations yield $\D/\H=(3.3\pm0.3)\times10^{-5}$ (PKS1937-1009),
$(4.0\pm0.7)\times10^{-5}$ (Q1009+2956), and
$(2.5\pm0.2)\times10^{-5}$ (HS0105+1619); their average value
\begin{equation}
\D/\H = (3.0\pm0.4)\times10^{-5}
\label{D}
\end{equation}
has been widely promoted as {\em the} primordial abundance
\cite{ome}. However the $\chi^2$ for this average is 7.1 implying
that systematic uncertainties have been underestimated, or that there
is real dispersion in the abundance. Other values have been reported
in different (damped Lyman-$\alpha$) systems which have a higher
column density of neutral \H, viz. $\D/\H=(2.24\pm0.67)\times10^{-5}$
(Q0347-3819) \cite{dod} and $\D/\H=(1.65\pm0.35)\times10^{-5}$
(Q2206-199) \cite{pb}. Moreover, allowing for a more complex velocity
structure than assumed in these analyses raises the inferred abundance
by upto $\sim50\%$ \cite{lev}. Even the ISM value of D/H now shows
unexpected scatter of a factor of 2 \cite{son}. All this may indicate
significant processing of the \D\ abundance even at high redshift, as
indicated in Fig.\ref{DHe3} \cite{foscv}. Given these uncertainties,we
conservatively bound the primordial abundance with an upper limit set
by the non-detection of D absorption in a high-redshift system
(Q0130-4021) \cite{ktblo}, and the lower limit set by the local
interstellar value \cite{lin}, both at $2\sigma$:
\begin{equation}
1.3 \times 10^{-5} < \D/\H|_\p < 9.7 \times 10^{-5} .
\label{Dp}
\end{equation}

For \he3, the only observations available are in the Solar system and
(high-metallicity) \hii\ regions in our Galaxy \cite{bbrw}. This makes
inference of the primordial abundance difficult, a problem compounded
by the fact that stellar nucleosynthesis models for \he3 are in
conflict with observations \cite{dst}. Such conflicts can perhaps be
resolved if a large fraction of low mass stars destroy \he3 by
internal mixing driven by stellar rotation, consistent with the
observed $^{12}$C/$^{13}$C ratios \cite{char}. The observed abundance
`plateau' in \hii\ regions (see Fig.\ref{DHe3}) then implies a
limit on the primordial value of $\he3/\H<(1.9\pm0.6)\times10^{-5}$
\cite{bbrw}, which is consistent with the other constraints we
discuss.

\subsection{The baryon density from standard (and non-standard) BBN} 

The observationally {\em inferred} light element abundances can now be used
to assess standard BBN for which the only free parameter is the
baryon-to-photon ratio $\eta$. (The implications of non-standard
physics for BBN will be considered shortly.) The overlap in the $\eta$
ranges spanned by the larger boxes in Fig.\ref{abundantlight}
indicates overall concordance. More quantitatively, accounting for
theoretical uncertainties as well as the statistical and systematic
errors in observations, there is acceptable agreement among the
abundances when \cite{pdg}
\begin{equation}
2.6 \le \eta_{10} \le 6.2 \ .
 \label{etarange}
\end{equation}
However the agreement is far less satisfactory if we use only the
quoted statistical errors in the observations. As seen in
Fig.\ref{abundantlight}, \he4 and \li7 are consistent with each other
but favor a value of $\eta$ which is lower by $\sim2\sigma$ from the
value $\eta_{10}=5.9\pm0.4$ indicated by the D abundance (\ref{D}).
Additional studies are required to clarify if this discrepancy is
real.

The broad concordance range (\ref{etarange}) provides a measure of the
baryon content of the universe. With $n_\gamma$ fixed by the present
CMB temperature $T_0=2.725\pm0.001$~K, the baryonic fraction of the
critical density is $\Omega_\B=\rho_\B/\rho_{\rm
crit}\simeq\eta_{10}h^{-2}/274$ (where
$h\equiv\,H_0/100$~km\,s$^{-1}$\,Mpc$^{-1}$ is the present Hubble
parameter), so that
\begin{equation}
0.0095 \le \Omega_\B h^2 \le 0.023 \ , 
\label{OmegaB}
\end{equation} 
For comparison, if the \D\ abundance in Eq.(\ref{D}) is indeed its
primordial value then the implied baryon density is at the upper end
of the (95\% c.l.) range quoted above:
\begin{equation}
\Omega_\B h^2 = 0.020 \pm 0.0015\ . 
\label{OmegaB'}
\end{equation} 
In either case since $\Omega_\B\ll1$, baryons cannot close the
universe. Furthermore, the cosmic density of (optically) luminous
matter is $\Omega_{\rm lum}\simeq0.0024h^{-1}$ \cite{fhp}, so that
$\Omega_\B\gg\Omega_{\rm lum}$; most baryons are optically dark,
probably in the form of a $\sim10^6$~K X-ray emitting intergalactic
medium \cite{co}. Finally, given that $\Omega_{\rm M}\gtrsim0.3$, we
infer that most matter in the universe is not only dark but also takes
some non-baryonic (more precisely, non-nucleonic) form.

The above limits hold for the standard BBN model and one can ask to
what extent they can be modified if plausible changes are made to the
model. For example if there is an initial excess of electron neutrinos
over antineutrinos then $n-p$ equilibrium is shifted in favour of less
neutrons, leading to less \he4. However the accompanying increase in
the relativistic energy density speeds up the expansion rate and
increases the $n/p$ ratio at freeze-out, leading to more \he4,
although this effect is smaller. For neutrinos of other flavours which
do not participate in nuclear reactions only the latter effect was
presumed to operate, allowing the possibility of balancing a small
chemical potential in $\nu_e$ by a much larger chemical potential in
$\nu_{\mu,\tau}$, and thus substantially enlarging the concordance
range of $\eta_{10}$ \cite{degen}. However the recent recognition from
Solar and atmospheric neutrino experiments that the different flavours
are maximally mixed no longer permits such a hierarchy of chemical
potentials \cite{nuosc}, thus ruling out this possible loophole.
Another possible change to standard BBN is to allow inhomogeneities in
the baryon distribution, created e.g. during the QCD (de)confinement
transition. If the characteristic inhomogeneity scale exceeds the
neutron diffusion scale during BBN, then increasing the average value
of $\eta$ increases the synthesised abundances such that the
observational limits essentially rule out such inhomogeneities.
However fluctuations in $\eta$ on smaller scales will result in
neutrons escaping from the high density regions leading to spatial
variations in the $n/p$ ratio which might allow the upper limit to
$\eta$ to be raised substantially \cite{inhom1}. Recent calculations
show that \D\ and \he4 can indeed be matched even when $\eta$ is
raised by a factor of $\sim2$ by suitably tuning the amplitude and
scale of the fluctuations, but this results in unacceptable
overproduction of \li7 \cite{inhom2}. A variant on the above
possibility is to allow for regions of antimatter which annihilate
during or even after BBN (possible if baryogenesis occurs e.g. through
the late decay of a coherently oscillating scalar field); however the
\li7 abundance again restricts the possibility of raising the limit on
$\eta$ substantially \cite{antimat}. Finally the synthesised
abundances can be altered if a relic massive particle decays during or
after BBN generating electromagnetic and hadronic showers in the
radiation-dominated plasma. Interestingly enough the processed yields
of \D, \he4 and \li7 can then be made to match the observations even
for a universe closed by baryons \cite{desh}, however the production
of \li6 is excessive and argues against this possibility. In summary
the conservative range of baryon density (\ref{OmegaB}) inferred from
standard BBN appears to be reasonably robust; although non-standard
possibilities cannot be definitively ruled out altogether, they have
at least been examined in some detail and tests devised to constrain
them. This contrasts with the rather simple-minded manner in which
parameters have been extracted from CMB observations as discussed
below.

\section{The baryon density from the CMB}

The BBN prediction for the cosmic baryon density can be tested through
precision measurements of CMB temperature fluctuations on angular
scales smaller than the horizon at last scattering \cite{hs}. The
amplitudes of the acoustic peaks in the CMB angular power spectrum
provide an independent measure of $\eta$ \cite{jkks}. Creation (or
even `destruction') of photons between BBN and CMB decoupling can
distort the CMB spectrum, so there cannot be a significant change in
$\eta$ between BBN and CMB decoupling \cite{bs}. Thus comparison of
the two measurements is a key test for the standard cosmology;
agreement would provide e.g. a superb probe of galactic chemical
evolution \cite{cfo2}, while disagreement would require revision of
the standard picture.

\begin{figure}[htb]
\centerline{
\epsfysize5cm\epsffile{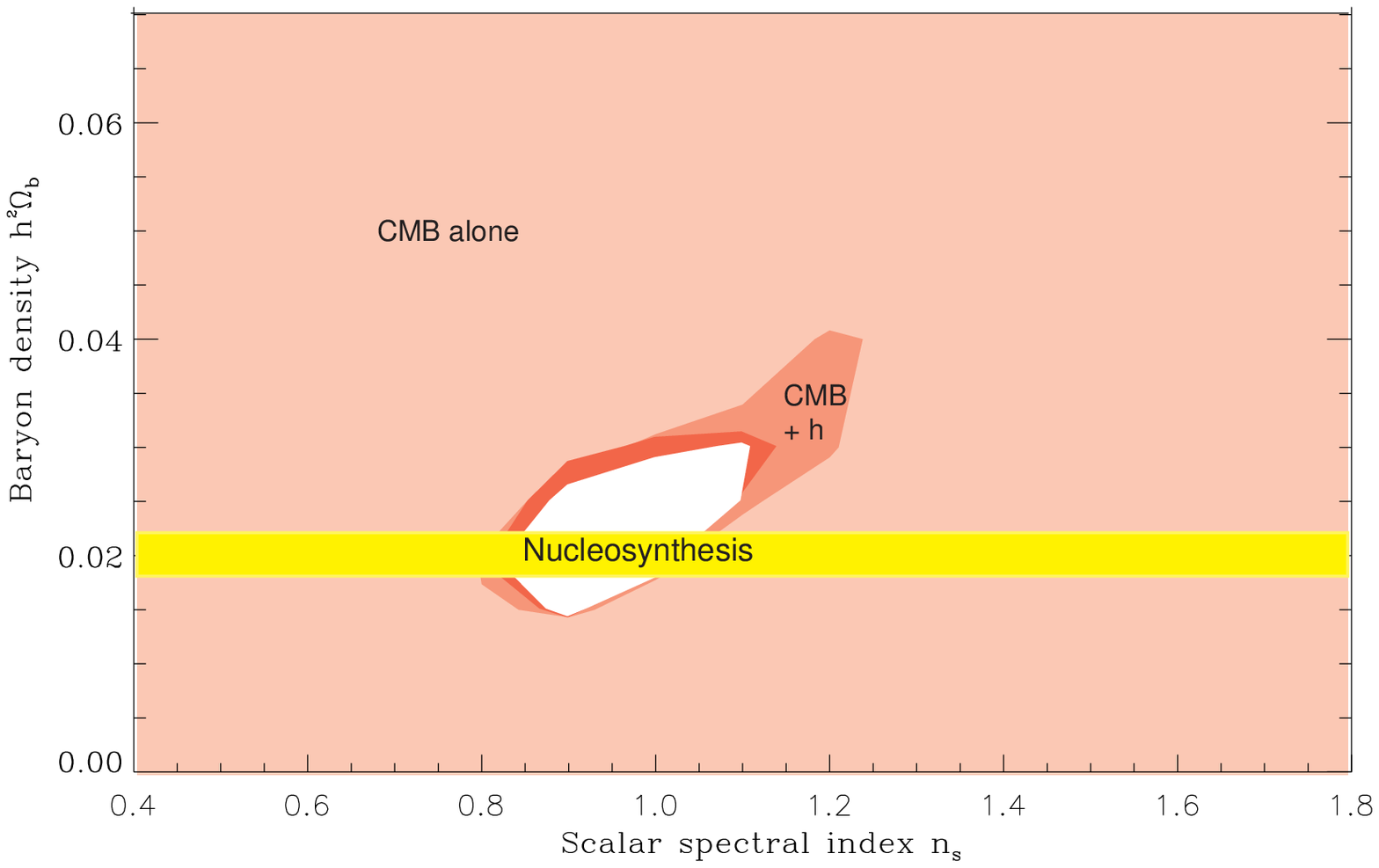}\hspace{1cm}
\epsfysize4.5cm\epsffile{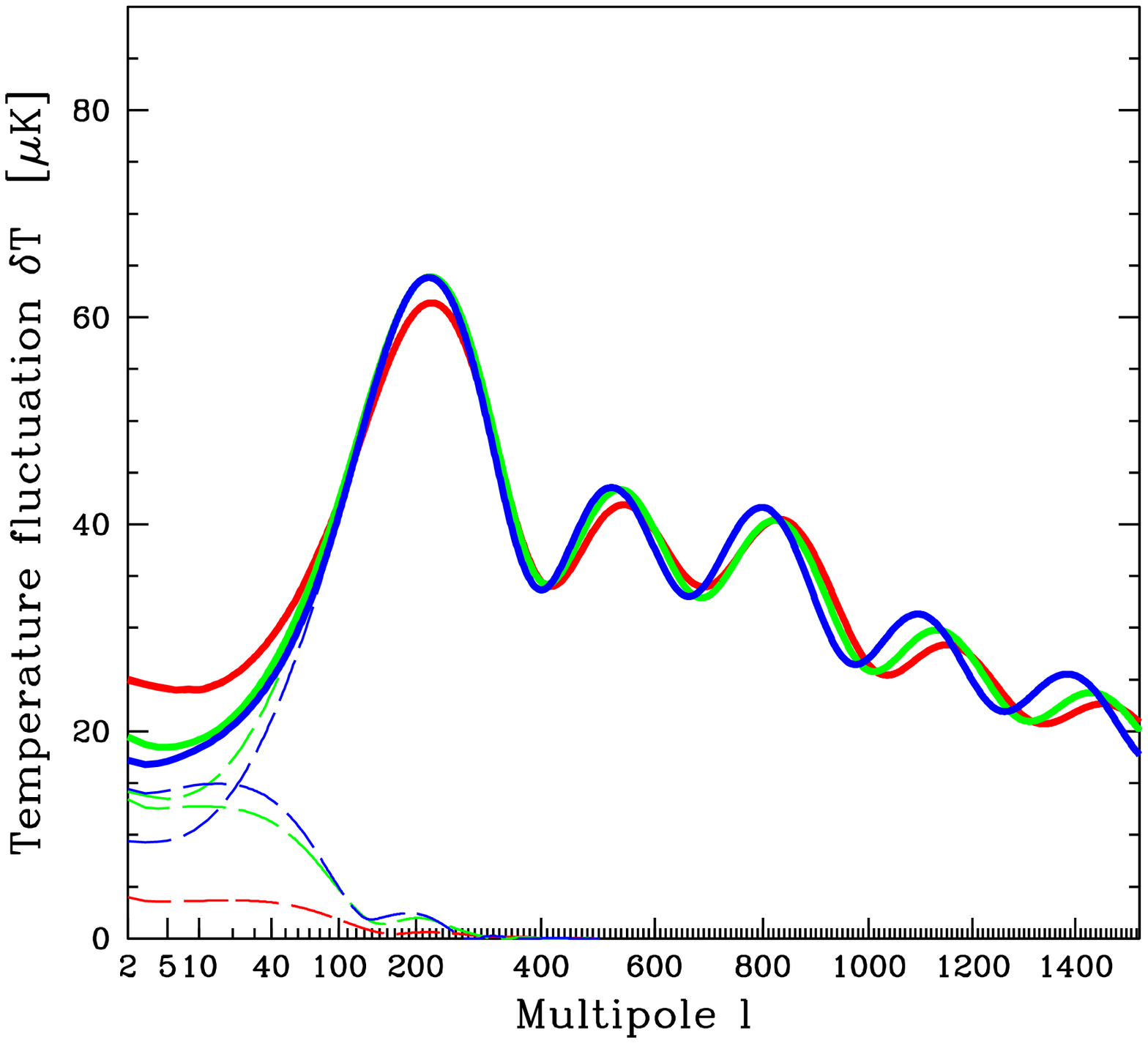}}
\bigskip
\caption{Fits to a compendium of CMB data (assuming adiabatic
primordial density perturbations) which demonstrate the inherent
`degeneracies' in parameter space. The left panel shows how the
inferred baryon density increases with the adopted (power-law)
spectral index. The right panel shows 3 models which fit the data
equally well but have different values of (from top to bottom)
$\Omega_{\B}h^2=$ 0.02 (red), 0.03 (green) and 0.04 (blue); the
increase in the baryon density is compensated by decreasing the amount
of dark matter, while increasing the dark energy content and the
spectral index (from Ref.\protect\cite{wtz}).}
\label{tilt}
\end{figure}

However as with other cosmological parameter determinations from CMB
data, the derived $\eta_\CMB$ depends on the adopted `priors', in
particular the assumed nature of primordial density perturbations The
standard expectation from simple inflationary models is for an
adiabatic perturbation with a spectrum that is close to the
`Harrison-Zeldovich' scale-invariant form \cite{infl}. However it is
perhaps not widely appreciated that there is no {\em physical} basis
for such `simple' models! In order to obtain the required extremely
flat potential for the inflaton it is essential to invoke supergravity
to protect against radiative corrections; in such models inflation
must occur far below the Planck scale and it is quite natural for the
spectrum to be significantly `tilted' below scale-invariance \cite{rs}
or even have sharp features at particular scales \cite{step}. As seen
in Fig.\ref{tilt}, the baryon density inferred from the CMB data is
sensitive to the assumed spectral slope $n_s$ even if a scale-free
spectrum is assumed; agreement with the BBN value (\ref{OmegaB}) does
{\em require} a significant tilt \cite{wtz}. Another way to have less
power on small scales in order to match the CMB data with the BBN
baryon density is by having a `step' in the spectrum at a scale of
$\sim50h^{-1}$~Mpc (as was suggested by an analysis of galaxy
clustering on the APM catalogue); this too compromises the ``precision
cosmology'' programme by dramatically altering the inferred
cosmological parameters such as the matter density and the
cosmological constant \cite{bgss}.

It is in this light that the much advertised \cite{mel} agreement
between the BBN baryon density (\ref{OmegaB'}) and the value
$\Omega_{\B}h^2=0.022^{+0.004}_{-0.003}$ inferred from the new
BOOMERanG \cite{boom} and DASI \cite{dasi} data, should be
evaluated.~\footnote{Note that MAXIMA-1 initially reported
$\Omega_{\B}h^2=0.033\pm0.006$ but a later (frequentist) analysis
finds $\Omega_{\B}h^2=0.026\pm^{+0.010}_{-0.006}$
\cite{maxima}. Observations at very small angles by the CBI suggest a
much smaller value $\Omega_{\B}h^2\sim0.009$ but the uncertainties are
large enough to allow consistency with the other results \cite{cbi}.}
Removal of the degeneracies referred to above will primarily require
{\em independent} measures of the primordial density perturbation
spectrum from studies of large-scale structure (LSS), as has been
attempted recently using data from the 2dFGRS survey
\cite{2df}. However such analyses still assume that the primordial
density perturbation is {\em scale-free} --- present data cannot
either confirm or rule out possible features in the spectrum
\cite{egl}, the presence of which can however significantly alter the
extracted cosmological parameters \cite{bgss} (if restrictive `priors'
are not imposed, e.g. on $h$). Moreover the normalisation of the
primordial scalar perturbation at large-scales to the COBE data
remains uncertain due to the possible contribution of gravitational
waves generated during inflation (shown as the decaying dashed lines
in Fig.\ref{tilt}).

\begin{figure}[htb]
\centerline{
\epsfxsize6.5cm\epsffile{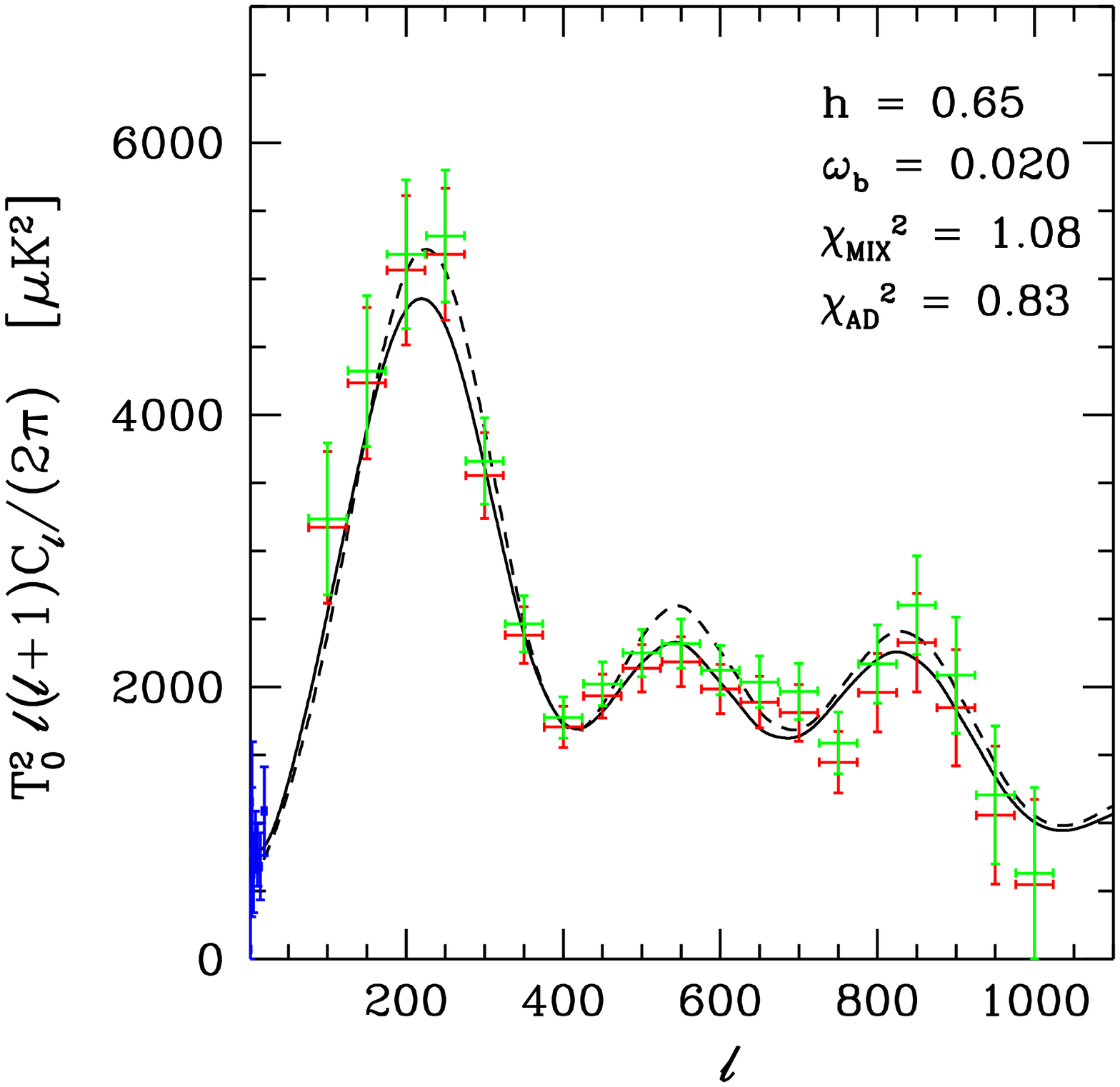}
\epsfxsize6.5cm\epsffile{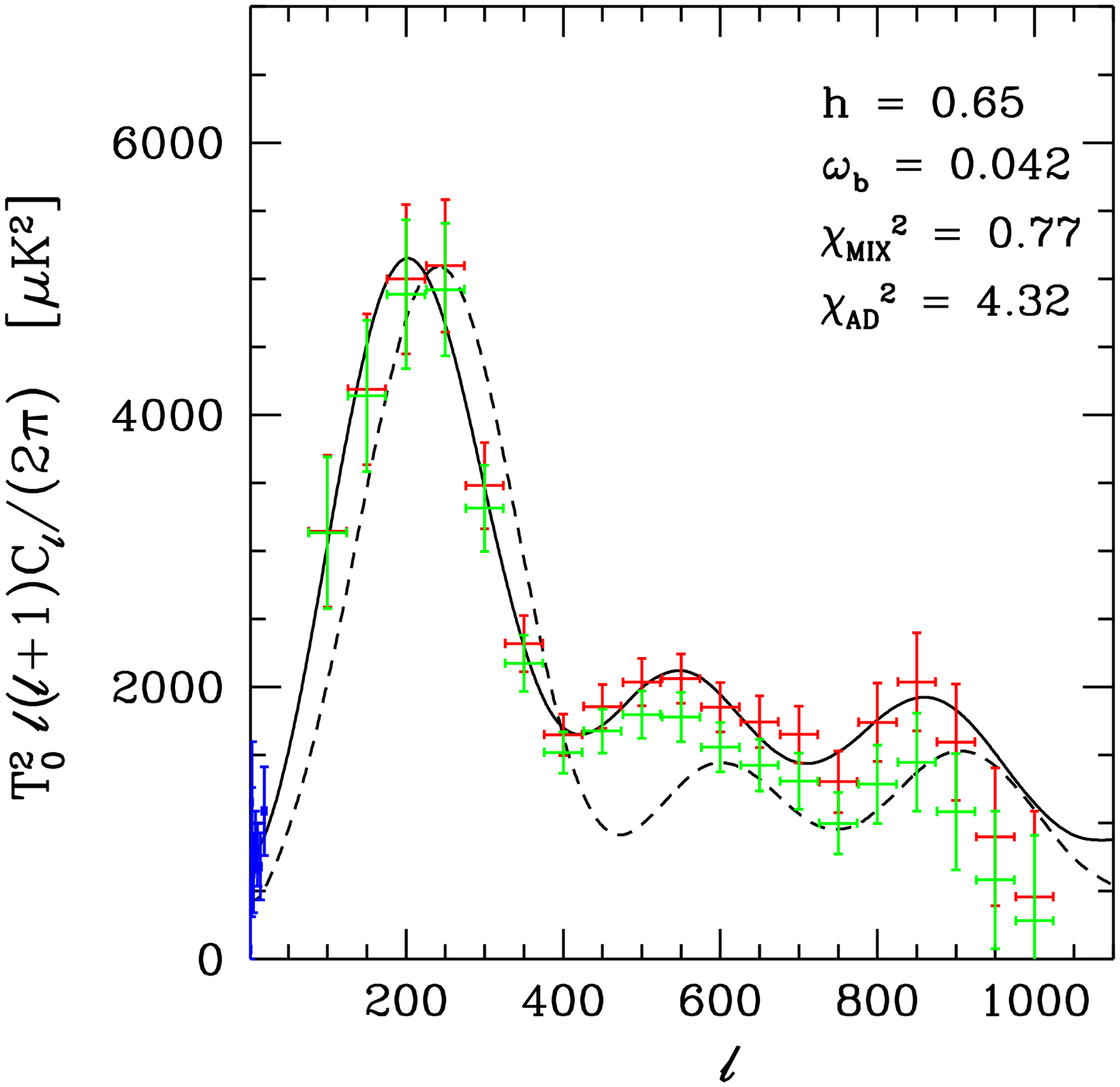}}
\bigskip
\caption{Fits to the CMB data assuming purely adiabatic perturbations
(dashed line) and mixed perturbations with $12\%$ isocurvature content
(solid line), adopting the baryon density
$\omega_\B=\Omega_{\B}h^2=0.02$ indicated by BBN (left panel), and a
value twice as large (right panel); note that the adiabatic case is
now a poor fit but the mixed case (with $69\%$ isocurvature component)
is quite acceptable (from Ref.\protect\cite{trd}).}
\label{iso}
\end{figure}

Even more dramatic changes to the values inferred from the CMB data
are possible if there are isocurvature modes present; the most general
cosmological perturbation can in fact contain four such modes --- in
baryons, dark matter and (two in) neutrinos \cite{bmt}. As shown in
Fig.\ref{iso}, a fit to the data {\em assuming} the BBN value of the
baryon density does not allow a significant admixture of isocurvature
modes; conversely if such modes are in fact dominant then the baryon
density required to fit the data is much larger \cite{trd}. To
distinguish experimentally between isocurvature and adiabatic
perturbations will require careful measurements of the CMB
polarisation which will be possible with the forthcoming PLANCK
surveyor \cite{pol}.

\section{Discussion}

It is clear that we are still some way away from establishing
rigorously that the baryon density inferred from CMB data equals the
value required by BBN. Nevertheless the fact that the two values are
within a factor $\sim2$ of each other is encouraging and has focussed
welcome attention on deeper issues. Foremost among these is our
fundamental ignorance concerning the nature and spectrum of the
primordial density perturbations. The prediction from `non-baroque'
models of inflation is of a purely adiabatic, nearly scale-invariant
perturbation. While the CMB and LSS data are, within present
uncertainties, consistent with this, it would be extremely surprising
if future precision data continued to support this naive model. The
physics of inflation lies beyond the Standard Model and general
expectations, while necessarily speculative, are of a much richer
phenomenology. Observationally, the breaking point may well come from
observations and modelling of the Lyman-$\alpha$ forest which yield an
independent measure of the baryon density \cite{lyalpha,lyalpha2}; a
recent study \cite{hui} finds a value of
$\Omega_{\B}h^2=0.045\pm0.008$ if the ionising UV background is as
intense as is indicated by observations of Lyman break galaxies
\cite{spa}. Further such measurements will soon provide a true
consistency check of the current paradigm. Perhaps one day we will
even have precision cosmology!

\section{Acknowledgements}

I am grateful to Ludwik Celnekier and Jean Tran Than Van for the
invitation to speak at this meeting, and to Brian Fields for allowing me
to draw on our joint review of BBN.

\end{document}